# Graph Neural Network-Accelerated Network-Reconfigured Optimal Power Flow

Thuan Pham, *Student Member, IEEE* and Xingpeng Li, *Senior Member, IEEE*

*Abstract*— Optimal power flow (OPF) has been used for real-time grid operations. Prior efforts demonstrated that utilizing flexibility from dynamic topologies will improve grid efficiency. However, this will convert the linear OPF into a mixed-integer linear programming network-reconfigured OPF (NR-OPF) problem, substantially increasing the computing time. Thus, a machine learning (ML)-based approach, particularly utilizing graph neural network (GNN), is proposed to accelerate the solution process. The GNN model is trained offline to predict the best topology before entering the optimization stage. In addition, this paper proposes an offline pre-ML filter layer to reduce GNN model size and training time while improving its accuracy. A fast online post-ML selection layer is also proposed to analyze GNN predictions and then select a subset of predicted NR solutions with high confidence. Case studies have demonstrated superior performance of the proposed GNN-accelerated NR-OPF method augmented with the proposed pre-ML and post-ML layers.

## I. INTRODUCTION

Transmission networks are highly meshed networks since they are critical electricity infrastructure that was built with redundancy. This indicates there are usually multiple electrical paths for long-distance power transfer from one area to another, enabling flexible transmission with dynamic network topologies. However, transmission networks have been traditionally considered static in real-time grid operations in the power system industry. For instance, optimal power flow (OPF) that is used to perform real-time generation redispatch does not model network reconfiguration. While traditional OPF only conducts generation-side regulation, the exploration of extra topology-aware regulation mechanisms in OPF will improve the grid efficiency and reliability.

Large power systems often face congestion, especially during peak times. Building new transmission lines is costly and time-consuming. Network-reconfigured optimal power flow (NR-OPF) is a solution that optimizes the existing network layout and power flows throughout the grid to make it more efficient. NR-OPF, also referred to as optimal transmission switching (OTS), can reduce network congestion, allowing for more renewable energy and lowering the total cost. It offers benefits like improved reliability, lower losses, and better voltage control [1]. Essentially, NR-OPF helps make the power grid more efficient and adaptable.

Thuan Pham and Xingpeng Li are with the Department of Electrical and Computer Engineering, University of Houston, Houston, TX, 77204, USA. (e-mail: tdpham7@cougarnet.uh.edu; xli83@central.uh.edu).

NR-OPF has been widely researched since its introduction. It can be applied to both short-term operations and long-term planning, at both the transmission and distribution level [2]. Importantly, it helps integrate large amounts of renewable energy into the grid. NR-OPF strategically changes the network layout by flexibly switching transmission lines on or off. This improves power flow distributions and makes better use of existing infrastructure. By identifying which lines to switch, grid operators can relieve lines congestion, improve voltage, and make the system more reliable [3].

NR-OPF can address suboptimal power system performance resulting from variability in load and generation patterns. Different scenarios of power generation and load profile may require distinct network topologies to achieve optimal efficiency. Consequently, a network optimized for one set of conditions may perform sub optimally when those conditions change [4]. Extending linear programming (LP) OPF problems, NR-OPF problems are commonly modeled as mixed-integer linear programming (MILP) problems. These formulations utilize binary variables and the big-M method for linearization, where the binary variables represent the switching status of each transmission line.

For large-scale networks with numerous transmission lines, the combinatorial explosion of binary variables significantly increases the computational complexity, leading to extended solution times. It is computationally demanding to determine the switching status of each transmission line in NR-OPF problems [5] [6]. Despite the substantial advantages NR-OPF offers over traditional OPF methods, these MILP-based NR-OPF problems pose significant challenges. The large and complex solution space makes it difficult to guarantee a globally optimal solution. The combinatorial nature of network topology—where the number of possible configurations grows exponentially with network size—further exacerbates the complexity of the problem [7].

The high computational complexity is a major bottleneck preventing the industrial adoption of the enhanced NR-OPF as a real-time application. Traditional optimization methods often struggle to handle this rapid increase in complexity, leading to potentially long solving times and suboptimal solutions. This inherent difficulty underscores the need for innovative and efficient approaches to tackle these challenging optimization problems in the power systems domain. Therefore, enhancing the computational efficiency of the NR-OPF problem is paramount. Reducing the computational complexity of NR-OPF problems can be achieved by strategically limiting the number of switchable lines considered.

In [8], a method is proposed to identify and rank the constraints that actively shape the OPF feasible region, based on their influence on the cost function. A heuristic algorithm is then developed that iteratively removes these constraints and solves a series of standard OPF problems, leading to more efficient solutions. In [9], different criteria concerning congestions, limit violation and production costs are used to identify candidate lines for switching. Machine learning (ML), specifically reinforcement learning and graph neural network (GNN), has been used in [10] to develop a heuristic algorithm for OTS at the distribution level. A new method for managing overloaded transmission lines during contingencies is offered in [11] to change the network's structure using NR-OPF. It introduces a two-stage optimization model that uses a DC model for an initial solution followed by an AC model for further optimization. Previous approaches haven't fully leveraged the potential of diverse system topologies or adapted to varying load profiles when identifying switching lines for transmission systems. There is a need for an adaptable model that can swiftly modify system layouts, leading to enhanced efficiency and reliability in power grids.

In this paper, we focused on utilizing ML, especially GNN, as an alternative method to speed up the NR-OPF solving process and address the drawbacks of existing works. Graph neural networks have shown promise in various power system applications such as fault detection [12], time-series prediction [13], and power flow calculation [14]. Researchers are actively investigating different GNN types to further enhance power system control and optimization. GNN was explored to reduce the number of constraints in base-case OPF problems [15]. For *N*-1 OPF, GNN was used in [16] to predict both base-case and contingency-case line congestions and remove non-congested lines in the monitoring set to reduce the model size and speed up the computing time. However, these methods are not applicable for NR-OPF, a much more complex MILP problem, due to the condition-based optimal dynamic topology scheme with NR leading to varying congestion bottlenecks. There were very limited studies on accelerating NR-OPF using ML. To bridge the gap, a novel GNN-accelerated NR-OPF (GaNR-OPF) approach with unique pre-ML filter and post-ML selection steps is developed in this paper to reduce NR-OPF model complexity, accelerate the solving process and thus reduce the computing time with quality solutions.

The contributions and innovations of this work are summarized as follows:
- First, we developed and trained a full GNN (FGNN) model to determine the optimal topology by predicting the switching status of each transmission line before entering the optimization stage of solving NR-OPF.
- Second, we developed an offline pre-ML filter step to identify non-critical lines whose switching status remains the same irrespective of the system loading levels, such that these non-critical lines will not be considered, leading to a reduced GNN (RGNN) model with less training time. The proposed RGNN model focuses the prediction process exclusively on the critical lines that influence the system's topology optimization decisions, streamline the utilization of ML technologies, and further improve its accuracy compared to the FGNN model.
- When adopting the predictions of FGNN or RGNN, it will convert the MILP NR-OPF into an LP OPF model with no more binary variables; however, it would result in increased cost or even infeasibility. To address this issue, we developed a fast online post-ML selection step to analyze the GNN predictions and then only select a subset of predicted NR solutions with high confidence for adoption. The selection process ensures that the final selected predictions are as reliable as possible, contributing to the overall robustness of the GaNR-OPF model. With this post-ML step, the binary variable representing the switching status of selected lines will be fixed before entering the optimization stage. The full NR-OPF is transformed into a reduced NR-OPF. The reduced NR-OPF will still be a MILP problem, but it will have much fewer binary variables, ensuring solution quality while still reducing the computing time substantially.
- To compare against the traditional full NR-OPF model, we have formulated a total of four methods using GaNR-OPF model: (i) FGNN-LP with neither the pre-ML layer nor the post-ML layer, (ii) FGNN-MILP with the post-ML layer, (iii) RGNN with the pre-ML layer, and (iv) the proposed RGNN-MILP with both the pre-ML and post-ML layers. Case studies have demonstrated RGNN-MILP method outperforms other benchmark methods. GaNR-OPF model, using RGNN-MILP, produces solutions with identical objective total costs to the full NR-OPF solutions and without any constraint violations. The proposed RGNN-MILP method demonstrated a substantial decrease in the computational time needed to find optimal solutions.

The remaining sections of this paper are organized as follows. Section II provides a comprehensive introduction to NR-OPF. Section III introduces GNN and explores the architectural layers. Section IV explains the methodologies employed in our research. Section V presents a detailed analysis of the simulation. In Section VI, we draw conclusions based on our findings, summarize the key takeaways and discuss the implications of our research.

II. NETWORK-RECONFIGURED OPTIMAL POWER FLOW

Network reconfiguration is a powerful concept incorporated into the OPF problem, empowering system operators with the ability to strategically and temporarily disconnect specific transmission lines from the network. This integration allows for the simultaneous optimization of both the grid's topology and the generation dispatch, ensuring a balance between operational efficiency and maintaining the system's overall reliability.

For the given traditional OPF problem below, with a bus set *N*, a branch set *K*, and a generator set *G*, the objective function is to minimize the total system generation cost as shown in (1), in which $P_g$ denotes the dispatched generation from each generator unit $g$ while $c_g$ denotes its associated operation cost. Constraints (2) and (3) set the generation reserve level, $P_g^{res}$, as 5% of the total amount of loads, $d_n$. In (4), generation output must meet generation limits, $P_g^{min}$ and $P_g^{max}$, as well as generation reserve. Line flows, $P_k$, are calculated in (5) based on the reactance of the line, $x_k$, and the angle of the bus ($\theta_{f(k)}$ and $\theta_{t(k)}$). Line flows are also subjected to line limit ratings, $RateA_k$, in (6). The nodal power balance for each node is enforced in (7).

$$\min \sum_n c_g P_g \qquad g \in G \qquad (1)$$
$$\sum_n P_g^{res} = 0.05 \times \sum_n d_n \qquad g \in G, n \in N \quad (2)$$
$$P_g^{res} \geq 0 \qquad g \in G \qquad (3)$$
$$P_g^{min} \leq P_g \leq P_g^{max} - P_g^{res} \qquad g \in G \qquad (4)$$
$$P_k = \frac{\theta_{f(k)} - \theta_{t(k)}}{x_k} \qquad k \in K \qquad (5)$$
$$-RateA_k \leq P_k \leq RateA_k \qquad k \in K \qquad (6)$$
$$\sum_n P_g + \sum_{n(f)} P_k + \sum_{n(t)} P_k = d_n \qquad g \in G, n \in N \quad (7)$$

In the NR-OPF problem, lines are selectively disconnected to improve the economic efficiency of the power system. Equation for line flows in (5) are replaced by (8) and (9). The variable $NR_k$ is a binary variable that describes the status of each line, where a value of 1 means ON and a value of 0 means OFF. Line limit ratings in (6) are multiplied with the variable $NR_k$ to produce (10) below.

The application of the big-M method in equations (8) and (9) facilitates linearization by introducing large negative constants associated with the constraints. These constants, denoted by *M*, are carefully chosen to ensure that they would not be part of any optimal solution if such a solution exists. This approach penalizes any violation of the constraints, effectively discouraging the optimization process from considering solutions that breach these limits.

$$-M(1 - NR_k) \leq P_k - \frac{\theta_{f(k)} - \theta_{t(k)}}{x_k} \qquad k \in K \qquad (8)$$
$$P_k - \frac{\theta_{f(k)} - \theta_{t(k)}}{x_k} \leq M(1 - NR_k) \qquad k \in K \qquad (9)$$
$$-NR_k \times RateA_k \leq P_k \leq RateA_k \times NR_k \quad k \in K \quad (10)$$

Moreover, this technique is commonly employed in formulating linear optimization problems where constraint violations incur a substantial positive penalty constant, again represented by *M*. The magnitude of *M* is crucial; it must be sufficiently large to render any constraint-violating solution significantly less attractive than feasible solutions [17] [18]. The big-M method plays a vital role in addressing non-linear constraints and ensuring constraint satisfaction within the optimization of the NR-OPF problem, thereby enhancing the effectiveness and reliability of the solution process.

In the context of NR-OPF, the NR vector transforms a linear programming problem into a more complex mixed-integer linear programming (MILP) problem. By strategically reducing the size of this NR vector—either through preprocessing techniques or by leveraging machine learning predictions—the overall complexity of the problem is significantly diminished. This simplification directly translates to faster computation times when seeking a solution.

## III. GRAPH NEURAL NETWORK

Machine learning algorithms, particularly convolutional neural networks (CNN) and deep neural networks (DNN), have been extensively utilized for power system optimization [19]. However, these models often neglect the inherent topological information within electrical networks, which could significantly enhance their learning process. GNN addresses this limitation by incorporating network topology through an adjacency matrix, enabling a more comprehensive understanding of node and edge features. This research gap will be bridged in this paper by developing a GNN-based acceleration approach to reduce the NR-OPF complexity and thus the computing time.

The unique capabilities of GNNs suggest they have great potential to enhance existing power system optimization tools, such as finding faster solutions for OPF. By effectively capturing the complex interdependencies within a power system network, GNNs could lead to more accurate predictions of important variables like load and generation. Ultimately, this could translate to faster and more efficient solutions for complex optimization problems in the power industry.

In electrical networks, buses are represented as nodes with associated generation and load profiles, while line flows and thermal limits are represented as edge features. The adjacency matrix defines the connectivity between buses. GNN leverages this global context to establish relationships among multiple node and edge features, facilitating optimizable transformations while preserving graph symmetries and permutation invariance.

GNNs have diverse applications in node-level, edge-level, and graph-level tasks. Graph-level tasks involve determining a single property for the entire graph, such as predicting properties of protein molecules based on their molecular structure. Edge-level tasks identify associations between nodes, as exemplified by analyzing citation networks to understand connections and quantify the value of edges (citations). By incorporating network topology, GNNs offer enhanced capabilities for power system optimization and various other domains where graph-structured data plays a crucial role.

GNN is specifically designed to work with network-structured data, as they can incorporate the network's topology into its input feature vectors. This allows them to understand and capture the dependencies and relationships that exist within complex graphs through a process called message passing between nodes. These message passing interactions allow information to flow across different layers of the GNN, making them especially effective for power system applications.

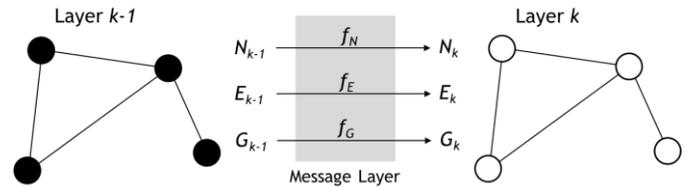

Fig. 1. Representation of GNN where node features N, edge features E, and graph features G are updated through a message layer.

During the GNN model's training phase, equation (15) illustrates how each node interacts with its neighbors in every forward pass. $h_i^k$ encapsulates all the feature information associated with the node $i$ for the current layer $k$, while $W_{self}^k$ and $W_{neigh}^k$ are weight matrices shared across all nodes, enabling the model to learn generalizable patterns from the network structure. $AGG(h_j^{k-1})$ combines feature information gathered from all neighboring nodes in previous layer, denoted by $\Omega_i$. In each forward pass, the model aggregates information from neighboring nodes and feeds it into the next training step using the activation function $\sigma$. With a well-structured GNN model comprising sufficient layers, feature information from the entire network can be harnessed to enhance the model's predictive capabilities. By exploiting the network's topology during training, we can gain valuable insights and boost the efficiency of power system optimization using GNNs.

In essence, equation (15) represents the core mechanism by which the GNN model learns to capture and propagate information throughout the power network, enabling it to make informed predictions and contribute to optimized power system operation.

$$h_i^k = \sigma(W_{self}^k \cdot h_i^{k-1} + W_{neigh}^k \cdot AGG(h_j^{k-1}, \forall j \in \Omega_i)) \qquad (15)$$

In this study, the XENet layer was employed as a graph convolutional layer within the proposed GNN model. Its selection was motivated by its capability to perform convolution operations on both edge and node tensors, as demonstrated in [20]. The XENet layer functions as a message-passing layer, concurrently considering both incoming and outgoing information from neighboring nodes. This ensures that a node's representation is influenced not only by the messages it receives but also by the messages it sends. The XENet architecture is mathematically described by the following equations.

$$f_{xy} = \varphi^{(f)}(p_x||p_y||q_{(x,y)}||q_{(y,x)}) \qquad (16)$$
$$f_x^{out} = \sum_{y \in N(x)} d^{out}(f_{xy}) \cdot f_{xy} \qquad (17)$$
$$f_x^{in} = \sum_{y \in N(x)} d^{in}(f_{xy}) \cdot f_{yx} \qquad (18)$$
$$p'_x = \varphi^{(n)}(p_x||f_x^{out}||f_x^{in}) \qquad (19)$$
$$q'_{(x,y)} = \varphi^{(q)}(f_{xy}) \qquad (20)$$

where $\varphi^{(f)}$, $\varphi^{(n)}$, $\varphi^{(q)}$ are multi-layer perceptrons with parametric rectified linear unit as the nonlinear activations, and $d^{out}$ and $d^{in}$ are two dense layers with sigmoid activations and a single scalar output.

For a given graph, vector $p_x$ contains the features associated with node $x$, and vector $q_{(x,y)}$ contains the features belonging to edge between node $x$ and node $y$. The XENet layer aggregates information of the feature stacks $f_{xy}$ in (16)-(18) for message-passing between each forward pass. By concatenating the node and edge attributes associated with the incoming and outgoing messages of each node in (16), the multi-layer perceptron $\varphi^{(f)}$ learns to process information such that a node's representation is based on the messages it receives as well as those it sends. The feature stacks are used to aggregate incoming/outgoing information, using self-attention to compute a weighted sum, in (17) and (18) respectively. The incoming/outcoming messages are concatenated and used to update the new node attributes, $q'_x$, of the graph in (19). Finally, some additional processing of the feature stacks through $\varphi^{(q)}$ in (20) allows us to compute new edge attributes, $q'_{(x,y)}$, that are dependent on the message exchange between nodes.

XENet plays a crucial role in our GNN model's architecture, allowing us to give special consideration to both incoming and outgoing edge attributes. By stacking features of nodes and edges and passing these learned features through each forward pass, it enables us to perform edge-level predictions across the network. Our proposed GaNR-OPF model relies on the XENet layer to accurately predict the switching status of each transmission line. It allows us to simplify the problem and create a computationally efficient NR-OPF model, thereby reducing the time required to find the optimal solution.

## IV. METHODOLOGIES

This section details the GaNR-OPF model, a novel approach for dynamic line switching status determination. The model is divided into two distinct stages: offline training stage, and online prediction stage.

To assess the effectiveness of the GaNR-OPF model, four benchmark methods are defined and implemented. Comparative analysis of these benchmarks will be conducted to evaluate the performance and advantages of the proposed method in terms of accuracy and efficiency.

### A. OFFLINE – Training Stage:

This stage involves the development and training of the underlying GNN model that will be used for prediction. The specifics of the training process, including data preparation, model selection, and optimization.

Initially, a substantial dataset was generated to simulate historical grid data, serving as the foundation for training the proposed GNN models. This involved running 10,000 full NR-OPF simulations on the IEEE 73-bus system, each with load profiles varying within ±7% of the initial profile. These simulations yielded the NR vector, indicating the ON/OFF status of each transmission line, which was then used as the training output for the GNN model.

As shown in Fig. 2, the data samples were partitioned into three distinct datasets: training, validation, and testing. The training dataset was specifically used to train the GNN model to predict line switching statuses. Both the training and validation datasets were instrumental in preparing the pre-ML filtering step. Predictions derived from the testing dataset were then evaluated in the post-ML selection step. Finally, a selected subset of lines was utilized to solve the NR-OPF problem, and the resulting solutions were rigorously checked for validity and constraint satisfaction.

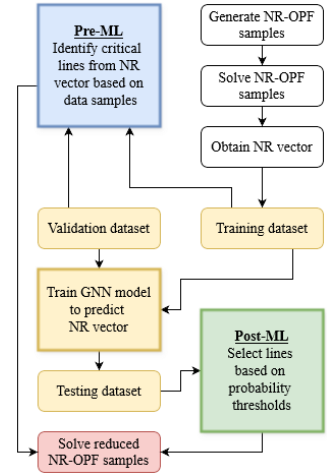

Fig. 2. Flowchart for the OFFLINE – Training Stage.

**Pre-ML Filtering Step**: The validation dataset was leveraged to analyze the NR vector and pinpoint "critical lines." These are defined as lines that consistently exhibit either a 100% ON or 100% OFF switching status across all the samples. Fig. 3 provides a visual representation of the line switching statuses derived from 10,000 NR-OPF solutions. A colormesh format is used where light (yellow) pixels signify an ON status and dark (purple) pixels indicate an OFF status.

By examining this graph, lines that remained either connected or disconnected throughout all the solutions were identified. These lines can be deemed non-critical as their status remains unaffected by varying loading conditions within the system. Consequently, these non-critical lines were removed from the GNN model's prediction process, resulting in a reduced GNN model. Strategic exclusion of these non-critical lines allows the GNN model to concentrate solely on predicting the switching status of the critical lines, those that are influenced by changes in loading conditions. This targeted approach not only streamlines the model's prediction task but also considerably improves its accuracy and efficiency by focusing on the most relevant lines within the IEEE 73-bus system.

reduced NR-OPF. This approach focuses the optimization process on the lines where the GNN model's predictions are less certain, potentially leading to a more efficient and accurate solution. With post-ML step, the NR-OPF problem would become a reduced NR-OPF problem, still MILP but much smaller in terms of complexities.

TABLE I
Examples of probabilities of prediction for a few lines in percentage (%)

| Line | 1 | 2 | 3 | 4 | 5 | 6 |
|---|---|---|---|---|---|---|
| OFF | 6.11% | 0.04% | 7.53% | 81.54% | 23.03% | 0.15% |
| ON | 93.89% | 99.96% | 92.47% | 18.46% | 76.97% | 99.85% |
| Post-ML Selection | Exclude | Select | Exclude | Exclude | Exclude | Select |

### B. ONLINE – Prediction Stage:

In this operational stage, the trained GNN model is deployed to analyze real-time data and predict the optimal switching status for each line. We deployed the following four different methods:

- **Benchmark FGNN-LP method**: uses only a full GNN model for prediction of switching status of all lines without any extra processing steps.
- **Benchmark FGNN-MILP method**: uses the post-ML selection step, after using a full GNN model for prediction of switching status of all lines, to convert selected line switching binary variables into parameters.
- **Benchmark RGNN-LP method**: uses the pre-ML filtering step to train a reduced GNN model for prediction of switching status of a subset of critical lines.
- **Proposed RGNN-MILP method**: uses both the pre-ML filtering step and post-ML selection step.

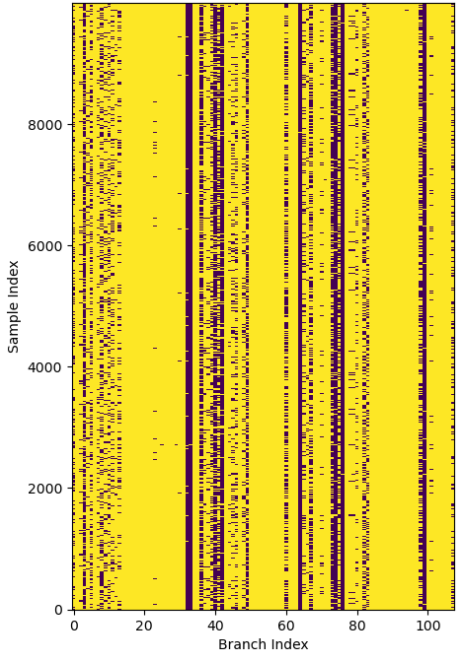

Fig. 3. Line switching status for 10,000 NR-OPF solutions data sample.

**Post-ML Selection Step**: The GNN model generated predictions for each line's switching status, expressed as probabilities of being ON or OFF. These probabilities summed to 100% for each line. To refine these predictions, a filter was applied: the lines with predicted probabilities between a lower limit of 5% and an upper limit of 95% were *Excluded*. These lines, in gray shading in TABLE I, represent those where the GNN model is less confident about their switching status. Lines with probabilities outside this range, indicating high confidence in either the ON or OFF state, were *Selected* - selected to have fixed status before entering the optimization stage, which enables to solve a reduced NR-OPF problem rather than the full NR-OPF problem. The switching status of the remaining *Excluded* lines is determined by solving the

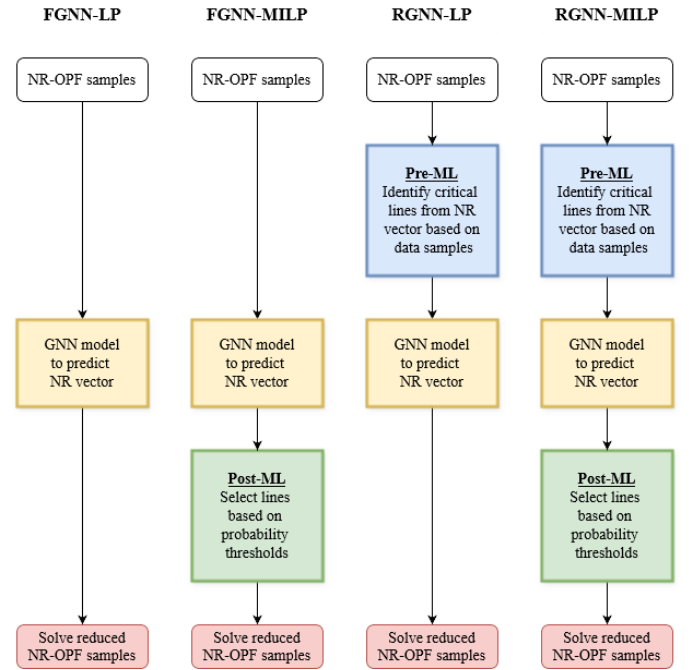

Fig. 4. Flowchart for the ONLINE – Prediction Stage using the four proposed benchmark methods.

Fig. 4 summarizes the flowchart for the four proposed methods. Once a smaller, more refined set of candidates for switching lines has been identified, these lines are then incorporated into the subsequent stage of the process: finding a solution for the reduced NR-OPF problem. By focusing on this smaller subset, the computational complexity of the NR-OPF problem is significantly reduced, potentially leading to faster and more efficient solutions.

## V. RESULTS AND ANALYSIS

GNN models and NR-OPF algorithm were written in Python. We used the Pyomo library [21] [22] and gurobi solver to solve NR-OPF problems. The GNN models were built using the spektral library [23]. All figures and graphs were generated using matplotlib [24] and networkx [25] library. A computer with 4 cores i7-4790K CPU, GTX3090 graphic card, and 32GB memory was utilized for all coding activities.

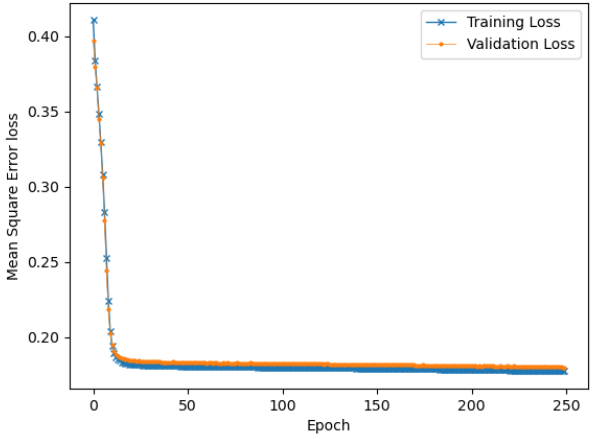

Fig. 5 MSE loss for training vs. validation of GNN model for the OFFLINE - training mode.

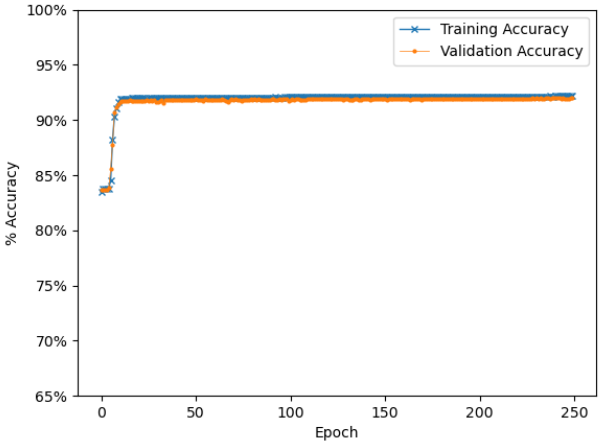

Fig. 6 Percent accuracy of prediction for training vs. validation of GNN model for the OFFLINE - training mode.

### A. Graph Neural Network

Fig. 5 and Fig. 6 demonstrate a successful training process for the GNN model. The close alignment between the training and validation loss curves suggests the model generalizes well to unseen data, minimizing the risk of overfitting. The smooth decline in loss, eventually reaching a plateau, signifies effective learning and convergence. Meanwhile, accuracy steadily improves, reaching around 93%. Together, these figures indicate that the training process achieved its goal: maximizing accuracy while minimizing loss.

### B. Performance Comparison on a Representative Sample

A representative sample was selected to illustrate the performance of the proposed RGNN-MILP method compared to other benchmark methods as detailed in TABLE II, while statistical results over 1,000 samples will be provided in the subsequent subsections. The data values are expressed in absolute values as well as in percentages relative to the complete NR-OPF solution. The ideal method would provide a solution exactly matching the full NR-OPF solution, as close to 100% as possible with significantly reduced solving time.

FGNN-LP and RGNN-LP demonstrate the fastest solving times compared to FGNN-MILP and RGNN-MILP. However, RGNN-LP shows the highest total cost among the methods evaluated, exceeding by up to 10%. FGNN-MILP has the longest solving time, approximately 84%. While FGNN-LP has the fastest solving time, it still has the second highest difference in total cost, over $2,000 in absolute value. RGNN-MILP is the only method that achieves a total cost nearly equivalent to the full NR-OPF problem while reducing solving time by as much as 98%.

TABLE II
Objective total cost and solving time of a representative sample

|  | Total Cost | | Solving Time | |
| --- | --- | --- | --- | --- |
|  | (%) | ($) | (%) | (second) |
| Full NR-OPF | 100 | 154,427.9 | 100 | 10.765 |
| FGNN-LP | 101.349 | 156,510.3 | 0.474 | 0.051 |
| FGNN-MILP | 99.998 | 154,424.9 | 84.742 | 9.122 |
| RGNN-LP | 110.079 | 169,992.7 | 0.520 | 0.056 |
| **RGNN-MILP** | 99.999 | 154,426.0 | 1.812 | 0.195 |

### C. Total Cost Statistics

TABLE III offers substantial evidence that the FGNN-MILP and RGNN-MILP methods consistently produce solutions with objective total cost that are highly accurate, and closely aligned with the actual solution. The metric, total cost in percent, is defined in equation (21). Both the median and the mean of these methods approach nearly 100%, indicating a reliable and precise performance. On the other hand, the FGNN-LP and RGNN-LP methods exhibit a much broader range of total costs, highlighting their variability in performance. The deviation from the actual solution in these methods can reach as high as 11%, indicating a less consistent and potentially less reliable approach. The significant difference in performance between the MILP-based methods and the LP-based methods underscores the superiority of the former in producing more accurate and stable results across different cases.

$$\% \, Total \, Cost = \frac{Total \, Cost \, using \, GaNR-OPF \, methods}{Total \, Cost \, using \, Full \, NR-OPF} \quad (21)$$

TABLE III
The total cost in percent for the four methods using equation (21)

|  | Total Cost in Percent (%) | | | | |
|---|---|---|---|---|---|
|  | Mean | Max | Min | Median | Std. Dev. |
| FGNN-LP | 102.17% | 103.72% | 100.29% | 102.23% | 0.80% |
| FGNN-MILP | 100.02% | 100.98% | 99.51% | 100.00% | 0.11% |
| RGNN-LP | 102.25% | 111.22% | 100.00% | 101.83% | 1.75% |
| **RGNN-MILP** | 100.05% | 100.56% | 99.92% | 100.02% | 0.09% |

The observation is further underscored by Fig. 7, which visually depicts the clustering of the majority of solutions near the 100% total cost mark. Among the various methods, the RGNN-MILP method stands out for its exceptional performance. Notably, more than 80% of the solutions solved using this method perfectly align with the total costs of the associated full NR-OPF solutions, effectively matching the optimal solution. The RGNN-MILP method's ability to consistently achieve results that closely mirror the actual cost of a full NR-OPF problem demonstrates its superior effectiveness in cost optimization.

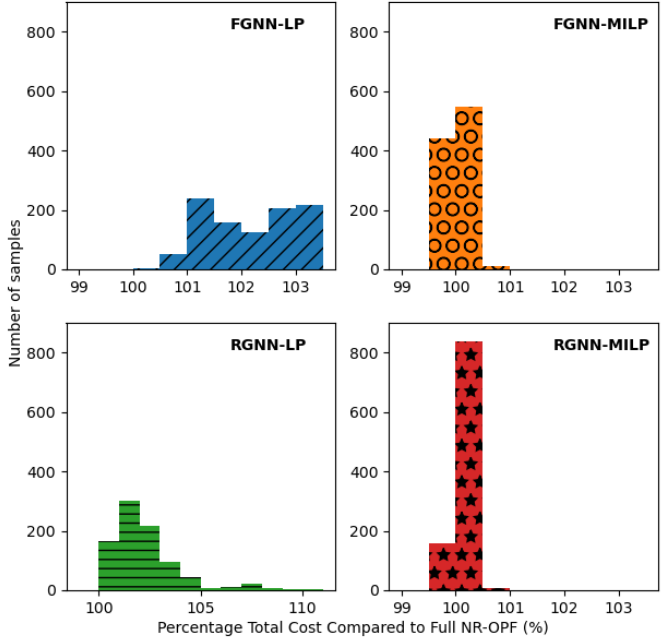

Fig. 7. Total cost of the four proposed benchmark methods.

### D. Solving Time Statistic

TABLE IV provides compelling evidence of remarkable reductions in solving time achieved for all four methods using the GaNR-OPF strategies. The FGNN-LP and RGNN-LP methods demonstrate particularly outstanding improvements, boasting mean solving times that are up to an astonishing 99% faster. The FGNN-MILP method also generally yields faster solving times, with an average reduction of around 40%. However, it is important to note that Fig. 8 reveals that certain samples processed by the FGNN-MILP method can take up to 10 times longer to solve compared to the original full NR-OPF problem. This suggests that while the FGNN-MILP method is generally faster, it may encounter specific instances where it is less efficient.

$$\% \, Solving \, Time = \frac{Solving \, Time \, using \, GaNR-OPF \, methods}{Solving \, Time \, using \, Full \, NR-OPF} \quad (22)$$

TABLE IV
Solving time in percent for the four methods using equation (22)

|  | Solving Time in Percent (%) | | | | |
|---|---|---|---|---|---|
|  | Mean | Max | Min | Median | Std. Dev. |
| FGNN-LP | 0.81% | 3.38% | 0.08% | 0.72% | 0.49% |
| FGNN-MILP | 60.23% | 1085.76% | 2.72% | 30.09% | 92.89% |
| RGNN-LP | 0.83% | 4.01% | 0.08% | 0.73% | 0.51% |
| **RGNN-MILP** | 6.34% | 83.77% | 0.41% | 3.47% | 8.24% |

In contrast, the proposed RGNN-MILP method, while not holding the absolute fastest solving time, still achieves impressive savings of 94% in solving times, with a median solving time of 3.5% for all samples. The consistent efficiency is highlighted, and the method underscores its potential as a valuable tool in reducing computational burden for NR-OPF problem.

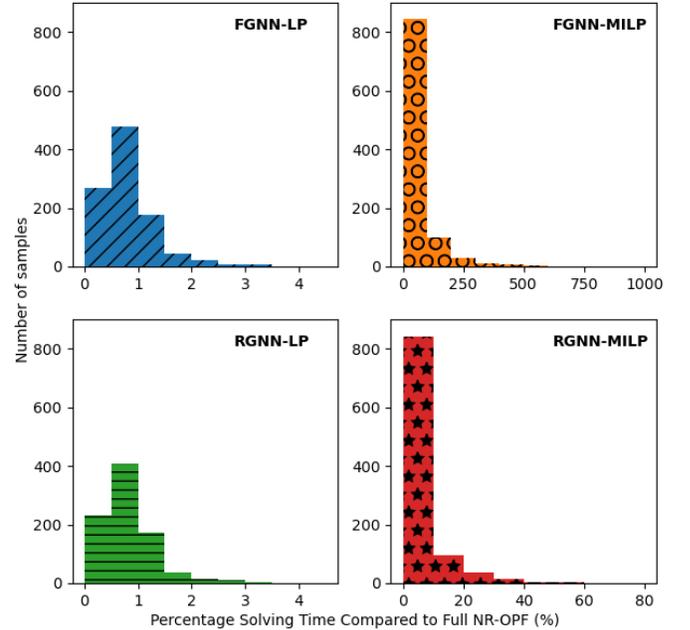

Fig. 8. Solving time of the four proposed benchmark methods.

### E. RGNN-MILP Method

Among the four benchmark methods, the proposed RGNN-MILP method emerges as a clear frontrunner for GaNR-OPF model in terms of overall speed and consistency. Its success can be attributed to the powerful combination of pre-ML and post-ML steps, which work in tandem to dramatically reduce the complexity of MILP problems. The reduction in complexity translates to an impressive 80% decrease in the number of critical lines required for the final solving NR-OPF step, streamlining the computational process significantly.

Remarkably, the RGNN-MILP method not only simplifies the problem but also maintains the quality of the solutions. It achieves solutions with a total cost that is virtually identical to the full NR-OPF solution, all while delivering a substantial 94% reduction in solving time on average. This exceptional performance highlights the RGNN-MILP method's unique ability to generate high-quality solutions without compromising computational efficiency.

The significant speedup is made possible by leveraging the predictive power of the GNN model and the synergistic interplay between the pre- and post-ML selection steps. By intelligently reducing the problem's complexity without sacrificing solution accuracy, the RGNN-MILP method offers a promising avenue for tackling computationally intensive tasks in a more efficient and effective manner.

## VI. Conclusion

The RGNN-MILP method demonstrates superior performance in addressing network reconfiguration challenges. Using this method for GaNR-OPF achieves near-identical objective total costs compared to the full NR-OPF problem while drastically reducing computational time by approximately 94%. The integration of pre-ML filtering and post-ML selection mechanisms proves crucial in achieving this optimal balance between solution speed and quality. While alternative methods may offer marginally faster solutions, they often do so at the expense of overall solution accuracy and optimality. GaNR-OPF has proven highly effective in managing a large-scale set of network reconfiguration problems efficiently. Although our current research is centered on a moderate-scale IEEE 73-bus system, extrapolation suggests comparable computational resource savings are achievable when scaling to larger systems comprising thousands of buses.

Further research will delve into alternative network reconfiguration strategies, including imposing constraints for line switching operations and utilizing virtual buses with distinct topologies for granular modeling of loads and generators. Additionally, the integration of GNNs into broader advanced OPF problem domains presents a promising avenue for future exploration.

## VII. References

[1] A. V. Ramesh and X. Li, "Security Constrained Unit Commitment with Corrective Transmission Switching," in *North American Power Symposium*, Wichita, 2019.

[2] H. Z. B. X. L. Rida Fatima, "Optimal Dynamic Reconfiguration of Distribution Networks," arXiv, [Online]. Available: https://arxiv.org/abs/2308.01984.

[3] S. R. Salkuti, "Congestion Management Using Optimal Transmission Switching," *IEEE Systems Journal,* vol. 12, no. 4, December 2018.

[4] E. B. Fisher, R. P. O'Neill and M. C. Ferris, "Optimal Transmission Switching," *IEEE Transactions on Power Systems*, vol. 23, no. 3, pp. 1346 - 1355, 2008.

[5] M. Flores, R. Romero and J. F. Franco, "An analysis of the optimal switching problem in transmission systems," in *ISGT Latin America*, Quito, 2017.

[6] M. Flores, L. H. Macedo and R. Romero, "Alternative Mathematical Models for the Optimal Transmission Switching Problem," *IEEE Systems Journal*, vol. 15, no. 1, pp. 1245 - 1255, 2021.

[7] M. Numan, M. F. Abbas, M. Yousif, S. S. M. Ghoneim, A. Mohammad and A. i. o. A. Noorwali, "The Role of Optimal Transmission Switching in Enhancing Grid Flexibility: A Review," *IEEE Access,* vol. 11, no. March, 2023.

[8] G. a. N. Sustainable Energy, "Feasible region-based heuristics for optimal transmission switching," *Sustainable Energy, Grids and Networks,* vol. 30, no. June, 2022.

[9] J. Wu and K. W. Cheung, "On selection of transmission line candidates for optimal transmission switching in large power networks," in *IEEE Power & Energy Society General Meeting*, Vancouver, 2013.

[10] T. Han and D. J. Hill, "Learning-Based Topology Optimization of Power Networks," *IEEE Transactions on Power Systems*, vol. 38, no. March, 2023.

[11] Y. Zhang, Z. Xu, J. Kuang, Y. Han, L. Zheng and R. Guo, "Corrective Power Network Reconfiguration for Eliminating Transmission Lines Overload," in *2020 IEEE 1st China International Youth Conference on Electrical Engineering*, Wuhan, 2020.

[12] F. Scarselli, M. Gori, A. C. Tsoi, M. Hagenbuchner and G. Monfardini, "The Graph Neural Network Model," *IEEE Transactions on Neural Networks,* vol. 20, no. 1, pp. 61 - 80, 2009.

[13] J. Zhou, G. Cui, S. Hu, Z. Zhang, C. Yang, Z. Liu, L. Wang, C. Li and M. Sun, "Graph neural networks: A review of methods and applications," *AI Open 2021,* vol. 1, pp. 57-81, 2020.

[14] W. Liao, B. Bak-Jensen, J. R. Pillai, Y. Wang and Y. Wang, "A Review of Graph Neural Networks and Their Applications in Power Systems," *Journal of Modern Power Systems and Clean Energy,* vol. 10, no. 2, pp. 345 - 360, March 2022.

[15] T. Pham and X. Li, "Reduced Optimal Power Flow Using Graph Neural Network," in *2022 North American Power Symposium (NAPS)*, Salt Lake City, 2022.

[16] T. Pham and X. Li, "N-1 Reduced Optimal Power Flow Using Augmented Hierarchical Graph Neural Network," [Online]. Available: https://arxiv.org/abs/2402.06226.

[17] L. D. Ramirez-Burgueno, Y. Sang and N. Santiago, "Improving the Computational Efficiency of Optimal Transmission Switching Problems," in *North American Power Symposium*, Salt Lake City, 2022.

[18] X. L. Qiushi Wang, "Evaluation of Battery Storage to Provide Virtual Transmission Service," arXiv, [Online]. Available: https://arxiv.org/abs/2309.07237.

[19] T. Pham and X. Li, "Neural Network-based Power Flow Model," in *2022 IEEE Green Technologies Conference*, Houston, 2022.

[20] J. B. Maguire, D. Grattarola, V. Mulligan, E. Klyshko and H. Melo, "XENet: Using a new graph convolution to accelerate the timeline for protein design on quantum computers," *PLoS Computational Biology,* no. September, 2021.

[21] W. E. Hart, J.-P. Watson and D. L. Woodruff, "Pyomo: modeling and solving mathematical programs in Python," *Mathematical Programming Computation,* 2011.

[22] W. E. Hart, C. D. Laird, J.-P. Watson, D. L. Woodruff, G. A. Hackebeil, B. L. Nicholson and J. D. Siirola, Pyomo – Optimization Modeling in Python, Vol. 67. Springer, 2017.

[23] D. Grattarola and C. Alippi, "Graph Neural Networks in TensorFlow and Keras with Spektral," *IEEE Computational Intelligence Magazine,* vol. 16, no. 1, pp. 99 - 106, Feb. 2021.

[24] J. D. Hunter, "Matplotlib: A 2D Graphics Environment," *Computing in Science & Engineering,* vol. 9, no. 3, pp. 90-95, 2007.

[25] A. A. Hagberg, D. A. Schult and P. J. Swart, "Exploring Network Structure, Dynamics, and Function using NetworkX," in *Proceedings of the 7th Python in Science Conference (SciPy2008)*, Pasadena, 2008.

[26] M. Tuo and X. Li, "Long-term Recurrent Convolutional Networks-based Inertia Estimation using Ambient Measurements," in *IEEE IAS Annual Meeting*, Detroit, MI, USA, 2022.

[27] V. Sridharan, M. Tuo and X. Li, "Wholesale Electricity Price Forecasting using Integrated Long-term Recurrent Convolutional Network Model," arXiv:2112.13681, Dec. 2021.

[28] P. Mandal, T. Senjyu, N. Urasaki, A. Yona, T. Funabashi and A. K. Srivastava, "Price Forecasting for Day-Ahead Electricity Market Using Recursive Neural Network," in *IEEE General Meeting Power& Energy Society*, Tampa, 2007.

[29] M. Zitnik, M. Agrawal and J. Leskovec, "Modeling polypharmacy side effects with graph convolutional networks," *Bioinformatics,* vol. 34, no. 13, p. i457–i466, 2018.

[30] M. Tuo, X. Li and T. Zhao, "Graph Neural Network-based Power Flow Model," in *55th North American Power Symposium*, Asheville, NC, 2023.

[31] R. Chen, J. Liu, F. Wang, H. Ren and Z. Zhen, "Graph Neural Network-Based Wind Farm Cluster Speed Prediction," in *SCEMS*, Jinan, 2020.